\documentclass[aps,pre,preprint,tightenlines,showpacs]{revtex4}
\usepackage{graphics}

\begin{document}
\title{Phaselocked patterns and amplitude death in a ring of delay coupled limit cycle oscillators}
\author{Ramana Dodla}
\email{ramana.dodla@nyu.edu}
\thanks{Present address: Center for Neural Science, New York University, New York, NY 10003. (The author's name 
appeared earlier as D. V. Ramana Reddy.)}
\author{Abhijit Sen}
\email{abhijit@ipr.res.in}
\author{George L. Johnston}
\email{glj@rcn.com}
\thanks{Present address: EduTron Corp., 5 Cox Road, Winchester, MA 01890.}
\affiliation{Institute for Plasma Research, Bhat, Gandhinagar 382428, India.}


\begin{abstract}
We study the existence and stability of phaselocked patterns and amplitude
death states in a closed chain of delay coupled identical limit cycle
oscillators that are near a supercritical Hopf bifurcation. The coupling is
limited to nearest neighbors and is linear. We analyze a model set of discrete
dynamical equations using the method of plane waves. The resultant dispersion
relation, which is valid for any arbitrary number of oscillators, displays
important differences from similar relations obtained from continuum models. We
discuss the general characteristics of the equilibrium states including their
dependencies on various system parameters. We next carry out a detailed linear
stability investigation of these states in order to delineate their actual
existence regions and to determine their parametric dependence on time delay.
Time delay is found to expand the range of possible phaselocked patterns and to
contribute favorably toward their stability. The amplitude death state is
studied in the parameter space of time delay and coupling strength. It is shown
that death island regions can exist for any number of oscillators $N$ in the
presence of finite time delay. A particularly interesting result is that the
size of an island is independent of $N$ when $N$ is even but is a decreasing
function of $N$ when $N$ is odd. 
\end{abstract}

\pacs{05.45.Xt, 87.10.+e}
\maketitle


\section{Introduction}
\label{SEC:intro}

The emergence of self-organized patterns is a common feature of nonequilibrium
systems undergoing phase transitions. Examples of such patterns abound in
nature ranging from intricate mosaic design on a butterfly wing,
vortex swirls in a turbulent stream to synchronous flashing of
fireflies \cite{Str:00}. The study of pattern formation and
self-synchronization using simple mathematical models consisting of
coupled differential equations has therefore been an active area of research
spanning many scientific disciplines. For example such models have
been employed to examine the collective output of arrays of lasers
\cite{Lam:64,VGEKL:97}, coupled magnetrons \cite{BSWSH:89}, Josephson-junctions
\cite{HBW:88b}, coupled chemical reactors \cite{CE:89,Kur:84book} and
electronic circuits \cite{BL:96}.  Rings of oscillators have also been useful
for modeling biological oscillations \cite{Win:80book,TTYNFE:01} and simulation
of phase relations between various animal gaits \cite{CS:94}. These simple
discrete models (usually composed of coupled limit cycle oscillators) are
amenable to some straightforward analysis and also direct numerical solutions
particularly when the number of oscillators is small. When the number of
oscillators becomes very large it is often convenient to go over to the
continuum limit of infinite oscillators and model the system with a partial
differential equation. Examples of the latter are reaction-diffusion type
mathematical equations, such as the time dependent complex Ginzburg-Landau
equation (CGLE), that have been widely used to simulate the dynamics of pattern
formation in fluids and other continuum systems \cite{CH:93}.  One of the most
popular of the discrete models is the `phase only model' that arose from the
pioneering contributions of Winfree \cite{Win:80book} and Kuramoto
\cite{Kur:76,KN:87}.  This model is valid as long as the coupling between the
oscillators is assumed to be weak so that amplitude variations can be
neglected. When the weak coupling approximation is relaxed and amplitude
variations are retained the system is found to admit new collective states such
as the amplitude death state \cite{Bar:85,AEK:90,Erm:90}, chaos and
multirhythmicity \cite{MS:90,MMS:91,OS:01,VV:02}. This strong coupling model has
received a great deal of attention in recent times
\cite{Erm:90,EK:90a,RE:00,SMM:92,CS:94}. The importance of time delay in
coupled oscillator systems and its effect on their collective dynamics has been
recognized for a long time. In particular there have been a large number of
studies devoted to the study of time delay coupled `phase only' 
models \cite{NSK:91,NTM:94,KPR:97,CW:98,Izh:98,EK:98,YS:99,BC:99,BC:99a,ES:03}. 
More recently the strong coupling model has also been generalized to include 
propagation time delays \cite{RSJ:98,RSJ:99,Atay1:03,Atay2:03}. A number of
interesting results regarding collective oscillations have emerged from this
generalized time delayed model some of which have been
experimentally verified \cite{HFRPO:00,RSJ:00exp,TFE:00}.

Our present work is devoted to further investigations and
understanding of the generalized time delay coupled model. Most of the past
investigations on this model have been restricted to collective states emerging
from {\it global coupling} among the oscillators, also known as the mean field
approximation. We have chosen to study here the equilibrium and stability of
collective modes emerging from a time delay coupled system of identical
oscillators in which the coupling is {\it local and restricted to nearest
neighbors}. The choice of this model is motivated by several considerations.
From a basic studies point of view the model offers us an opportunity to
compare and contrast the properties of the collective states of the local
coupling geometry to that of the globally coupled states. The model further
permits a parametric study of the influence of the number of oscillators $N$ on
the stability and existence properties of the equilibrium states. In the limit
of large $N$ the model also has close resemblance to the continuum CGLE and it
is of interest to delineate the important differences in the collective
properties of the two systems.
From a practical point of view our model has great relevance for many physical
systems including coupled magnetrons, study of collective phenomena in
excitable media, coupled laser systems, and small-world networks. For example,
an interesting study on the role of the geometry of the coupled magnetrons on
their rate of phaselocking was made by Benford et al. \cite{BSWSH:89}
who experimentally demonstrated production of higher microwave power at Giga
Watt levels through phaselocking. Our analysis of nearest neighbor coupled
oscillator array could provide important clues on not only the effect of
circular geometry but also on the effects of time delay in such a geometry.
A ring of coupled oscillators is also commonly encountered in a variety of
biological clocks \cite{Win:80book}. For
simplicity of analysis most mathematical studies of these rings have been
confined to an examination of their phase evolutions. Our model analysis can
provide  an extended understanding of these biological clocks with the
inclusion of time delays and amplitude effects.
Another interesting and more recent potential application of our work is in the
rapidly developing area of small-world networks where the simplest studied
configuration is often a ring of oscillators with nearest neighbor connections
dominating \cite{Str:01}. A recent study has explored the
role of time delays in a ring of connected oscillators with a special focus on
small-world networks \cite{ES:03}. We believe that our present study of
the ring configuration could help provide further insights in this direction
with the incorporation of amplitude effects.

The existence of
equilibrium states of locally coupled oscillator systems has been studied in
the past using group theoretic methods \cite{BCS:97,GS:86}. We use a plane
wave method and investigate both the existence and stability regions of the
equilibrium states of the coupled system. We derive a general dispersion
relation that can be used to determine the equilibrium states of any number of
coupled oscillators.  The method is further extended to a linear stability
analysis of these equilibrium states. Using this technique we delineate the
existence regions of the various equilibrium phase-locked patterns of the
coupled system. In the limit of small $N$ and in the absence of time delay our
results agree with past findings, but we also find new equilibrium states that
have not been noticed before. With time delay the existence regions change in a
significant manner and in some of the regimes we observe multirhythmicity.
Finally we also examine the nature of the death state in our system in the
presence of time delay.  Although time delay induced death state in locally
coupled identical oscillators has been numerically observed in one of our past
studies \cite{RSJ:98} there has been no systematic analytic investigation of
this phenomenon. We carry out such an analysis here and delineate the existence
of the death region as a function of the time delay and coupling strength
parameters. Our analysis yields an interesting result in that the size of the
{\it death island} is found to be independent of $N$ (the number of
oscillators) when $N$ is even but is a decreasing function of $N$ when $N$ is
odd. 

The paper is organized as follows. In the next section we present the model
equations and briefly discuss their similarities and differences with the
continuum model CGL equation.  Using the plane wave method we next derive the
dispersion relation in Section \ref{SEC:disp} and proceed to use it to delineate the
possible existence regions of the phase-locked patterns both in the presence
and absence of time delay. Section \ref{SEC:stab} is devoted to a linear stability analysis
of these states in order to identify their actual existence domain. In Section \ref{SEC:death}
we briefly look at the stability of the origin, that is the existence domain
of the death state, as a function of the time delay parameter and the coupling
strength. Our results are summarized and their implications discussed in the
concluding Section \ref{SEC:con}.


\section{Model equations}
\label{SEC:model}

We consider a one dimensional closed chain of delay coupled identical 
limit cycle oscillators that are close to a supercritical Hopf bifurcation.
Assuming the coupling to be linear
and of the nearest neighbor kind, we can write down the following set of model
equations to describe the time evolution of the oscillator states,
\begin{equation} 
\frac{\partial \psi_{j}}{\partial t} = (1 + i \omega_0 -\mid
\psi_{j} \mid^{2})\psi_{j} + K [ \psi_{j+1}(t-\tau) - \psi_{j}(t)
] + K [ \psi_{j-1}(t-\tau) - \psi_{j}(t) ],
\label{EQN:ring}
\end{equation}
$j=1,\ldots, N$, where $\psi_j (= X_j+i Y_j)$ is the complex amplitude of the
$j^{th}$ oscillator, $K > 0$ is the coupling strength, $\tau \ge 0$ is a fixed
time delay, and $\omega_0$ is the natural frequency of the uncoupled
oscillators.  The basic nonlinear oscillator we have chosen here is simply the
normal form of the supercritical Hopf bifurcation equation truncated to the
third order.  It is also known as the Stuart-Landau oscillator and has been
extensively used in the past as a model equation for studying nonlinear
phenomena in fluids, lasers, and Josephson junctions. For example, in the
absence of time delay ($\tau=0$), the above set of equations~(\ref{EQN:ring})
have been used to study the coupled wakes arising from a flow behind an array
of equally spaced parallel cylinders \cite{Fullana:97}.  Another interesting
connection to past work is the continuum limit for $\tau=0$, where one can
introduce a lattice spacing $a$, set $\psi_j=\psi(ja)$ in (\ref{EQN:ring}),
rescale $K$ as $K/a^{2}$ and then let $a \rightarrow 0$ with $ja \rightarrow
x$.  This reduces Eqs.~(\ref{EQN:ring}) for $\tau=0$ to the well known  complex
Ginzburg Landau equation,
\begin{equation}
\label{cgle}
\frac{\partial \psi(x,t)}{\partial t} = (1 +i\omega_0 - \left|\psi(x,t)\right|^2)\psi(x,t) + K\frac{\partial \psi^{2}(x,t)}{\partial x^{2}}.
\end{equation}
The CGLE has been extensively investigated in the past for its rich 
equilibrium states and for their applications to a variety of physical
situations \cite{MC:97,RLL:00}. In the presence of time delay it is not
meaningful to take the continuum limit since $a \rightarrow 0$ would also make
the propagation time tend to zero. One can instead take the so called
thermodynamic limit by letting $N \rightarrow \infty$ and the chain length $L$
$\rightarrow \infty$, while keeping $a=L/N$ fixed. In such a limit the time
delay parameter remains finite and meaningful.  The thermodynamic limit also
preserves the discrete (in space) character of the evolution equations and
therefore the corresponding equilibrium states can differ in character from
those obtained from a continuum limit. We will discuss some of these
differences in the next section where we derive and analyze the dispersion
relation for plane waves for our discrete set of equations~(\ref{EQN:ring}).


\section{Dispersion relation for plane waves}
\label{SEC:disp}

We seek plane wave solutions of Eq.~(\ref{EQN:ring}) of the form,
\begin{equation}
\Psi_{j} = R e^{i(jka+\omega t)},
\label{EQN:plane}
\end{equation}
where $a$ is the distance between any two adjacent oscillators 
and $k$ is the wave number such
that $-\pi \leq ka \leq \pi $. The values of $ka$, which define the phase
difference between adjoining oscillators, are further constrained by the
periodicity conditions inherent in a closed chain configuration. Since
$\psi_{N+1}$ must be identical to $\psi_1$ and $\psi_{0}$ must be identical to
$\psi_{N}$, we must satisfy the condition $\mathrm{e}^{i N ka} = 1$. This
implies that $ N ka = 2 m \pi$, $m = 0, 1, \ldots, N-1$, that is,
\begin{equation}
ka = m \frac{2\pi}{N}, ~~~m=0, 1, \ldots, N-1.
\label{spect}
\end{equation}
These discrete values of $ka$ are one of the defining properties of
the various phaselocked states of the coupled set of oscillators. Further 
characteristics of these states are described by a dispersion relation which
can be obtained by 
substituting (\ref{EQN:plane}) in Eq.~(\ref{EQN:ring}) to give,
$i\omega = 1+i\omega_0 -R^2 + 2K\left [\cos(ka)e^{-i\omega \tau} - 1 \right ].$
This leads to
\begin{eqnarray}
\label{EQN:freq}
\omega & = & \omega_0 - 2 K \sin(\omega \tau) \cos(ka),  \\
\label{EQN:amp}
R^2    & = & 1 - 2K + 2 K \cos(\omega \tau) \cos(ka).
\end{eqnarray}
Equations (\ref{EQN:freq}) and (\ref{EQN:amp}) which constitute the dispersion
relation for plane waves of our model equations can be compared to the
corresponding dispersion relation obtained from the CGLE, where we can
substitute $\psi(x,t) = R_{CGL} \exp(ikx-i\omega_{CGL} t)$ in (\ref{cgle}) to
get,
\begin{eqnarray}
\label{CGLE:disp1}
\omega_{CGL} & = & \omega_0, \\
\label{CGLE:disp2}
R_{CGL}^2    & = & 1 - K(ka)^2
\end{eqnarray}
where we have replaced $K$ by its scaled value.  Equations (\ref{CGLE:disp1})
and (\ref{CGLE:disp2}) can also be recovered from (\ref{EQN:freq}) and
(\ref{EQN:amp}) by setting $\tau=0$ and taking the long wavelength limit of $ka
\ll 1$. As we can see there are significant differences in the two dispersion
relations. While (\ref{EQN:freq}) and (\ref{EQN:amp}) are valid for any
arbitrary value of $N$, (\ref{CGLE:disp1}) and (\ref{CGLE:disp2}) are strictly
valid only in the limit of $N \rightarrow \infty$. The spectrum of $ka$ values
are therefore continuous for the latter whereas in the former they assume
discrete values that are determined by the magnitude of $N$.  For $\tau=0$,
Eqs.~(\ref{EQN:freq}) and (\ref{CGLE:disp1}) become identical, but
Eq.~(\ref{EQN:amp}) reduces to, 
\begin{equation}
R^2    =  1 - 2K + 2K \cos(ka) = 1 - 4K \sin^2(ka/2).
\label{disp_nod}
\end{equation}
Using $R^2 >0$ as the defining condition for the possibility of having a
phaselocked state, we see that this region is considerably reduced in the
case of the CGLE as compared to the discrete model equations for any given value
of the coupling parameter $K$. For example, at $K=1/4$, the discrete model allows
all modes from $0$ to $\pi$ to exist, whereas the continuum model has an upper
cutoff at $ka =2.0$. For $K>1/4$, cutoff regions also appear for the discrete
model and are given by the expression,
$$
f_1 = \cos^{-1}(1-1/2K) < ka < f_2 = 2\pi-\cos^{-1}(1-1/2K)
$$
As can be seen from the above expression, for $K>1/4$
the anti-phase locked state ($ka=\pi$) is now no longer a permitted state.
With the introduction of time delay, the existence region is not only
a function of $ka$ (for a given value of $K$) but also depends on 
$\omega$ which satisfies the transcendental equation (\ref{EQN:freq}).
This brings about a
significant modification of the existence domains and raises the 
interesting possibility of enabling many of the previously forbidden modes to 
exist. Since the dispersion relation now has a transcendental character
it also introduces additional branches of collective oscillations. 
To illustrate some of these modifications we consider a few simple examples.
Let $C = \cos(\omega\tau)$.
By choosing a value for $C$ and in turn defining $S=\sin(\omega\tau)$, we 
can immediately get an expression for $\omega$, and the 
resultant equation for $R^2$ defines the stability regions.

\begin{figure}
\centerline{\scalebox{0.8}{\includegraphics{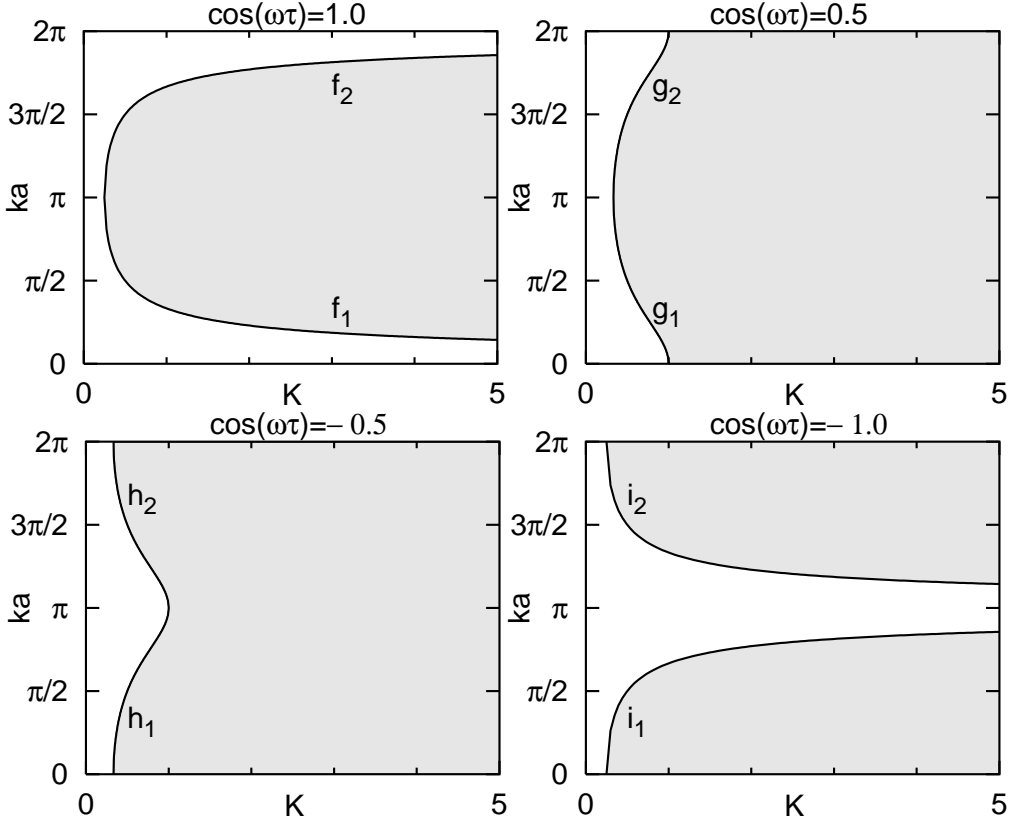}}}
\caption{Allowed (unshaded) and forbidden (shaded) wave modes in the presence of 
time delay for various values of $\cos(\omega\tau)$
\label{FIG:forbiddentau}}
\end{figure}

{\it Case (i):} First let $\cos(\omega\tau) = 1.$
This gives $\omega = \omega_0$, and
the square of the amplitude as $A_1 = 1 - 2 K + 2 K \cos(ka)$.
This also leads us to $\tau = 2 n \pi / \omega_0,$ $n=0, 1, 2, \ldots$.
This case also
includes the special case of $\tau=0$. Since $A_1 > 0$ for the existence
of the solutions, it is easily seen that 
the bounding region in $(K,ka)$ plane is defined by
$$f_1 < ka < f_2 = 2\pi-f_1$$
where $f_1 = \cos^{-1}(1-1/2K)$ is the forbidden region. 
Since $f_1$ does not intersect 
with the $K$ axis at any finite $K$, the region exists for all
$K>1/4$. In particular the in-phase state exists for all $K>0$, and the
anti-phase locked state ($ka=\pi$) no longer exists
if $K>1/4$. This forbidden region is illustrated in Fig.~\ref{FIG:forbiddentau}(a).

{\it Case (ii):} Let $\cos(\omega\tau)=1/2$. This gives the dispersion relation
$\omega=\omega_0 - \sqrt{3} \cos(ka)$. The corresponding delays
are defined by $\tau = (2n \pm 1/3)\pi/[\omega_0 - \sqrt{3} K \cos(ka)], 
n = 0, 1, 2, \ldots.$ The square of the amplitude is given by
$A_2 = 1 - 2K + K \cos(ka)$. Let $g_1 = \cos^{-1}(2-1/K)$.
The curve $g_1$ assumes values for $K$ between 1/3 and 1.
If $K>1$, $A_2<0$ for any $ka$. So all the modes are forbidden if $K>1$,
and the region between $K=1/3$ and $K=1$ as defined by
$$g_1 < ka < g_2 = 2\pi-g_1. $$
is also forbidden. In particular the in-phase solutions do not exist
if $K>1$ and the anti-phase solutions do not exist if $K>1/3$.
This region is illustrated in Fig.~\ref{FIG:forbiddentau}(b).

{\it Case (iii):} Let $\cos(\omega\tau)=-1/2$. This gives the dispersion relation
$\omega=\omega_0 + \sqrt{3} \cos(ka)$. The corresponding delays
are defined by $\tau = (2n \pm 1/3)\pi/[\omega_0 + \sqrt{3} K \cos(ka)], 
n = 0, 1, 2, \ldots.$ The square of the amplitude is given by
$A_3 = 1 - 2K - K \cos(ka)$. Let $h_1=\cos^{-1}(1/K-2)$.
If $K>1$, $A_3<0$ for all $ka$. So $K>1$ is the forbidden region.
Between $K=1/3$ and $K=1$ the region defined by
$$h_1 > ka > h_2 = 2\pi-h_1$$
is forbidden. The inequality signs are reversed in this case because 
the curvature of $h_1$ is different from that of $g_1$ or $f_1$.
In particular the in-phase solutions are forbidden if $K>1/3$ and the
anti-phase solutions are forbidden if $K>1$.
This region is illustrated in Fig.~\ref{FIG:forbiddentau}(c).

{\it Case (iv):} Let $\cos(\omega\tau)=-1$. This results in
the dispersion relation $\omega=\omega_0$ just as in case (i). 
This case corresponds to
$\tau = (2n\pi+1)\pi/\omega_0, n = 0, 1, 2, \ldots.$
The square of the amplitude is given by $A_4 = 1 - 2 K - 2 K \cos(ka)$.
Since $A_4=1$ at $ka=\pi$, in contrast to the three previous cases, 
the anti-phase locked solutions exist for all
values of $K$. At $ka=0$, $A_4=1-4K$. So the in-phase solutions are forbidden
for all $K>1/4$. The forbidden regions when $K>1/4$ are defined by
$$i_1 > ka > i_2 = 2\pi-i_1,$$ where $i_1 = \cos^{-1}(1/2K-1)$.
This region is illustrated in Fig.~\ref{FIG:forbiddentau}(d).
In fact we can derive a general expression for the existence
curves by choosing any arbitrary value for $C$. This in turn defines
the frequencies as $\omega = \omega_0 - 2 K S \cos(ka).$
After some simple algebra, the corresponding delays are given as
$$\tau = ( 2 n \pi + \cos^{-1}C ) / [ \omega_0 - 2 K S \cos(ka) ], ~~\mathrm{if}~~ S\ge 0,$$
$$\tau = ( 2 n \pi - \cos^{-1}C ) / [ \omega_0 - 2 K S \cos(ka) ], ~~\mathrm{if}~~ S < 0.$$
The curve that defines the boundary of the forbidden region is given by
$j_1 = \cos^{-1}(1/S-1/2KS)$ when $S\ne 0$. The case of $S=0$ is the same as
case (i) studied above.
\begin{figure}
\centerline{\scalebox{0.8}{\includegraphics{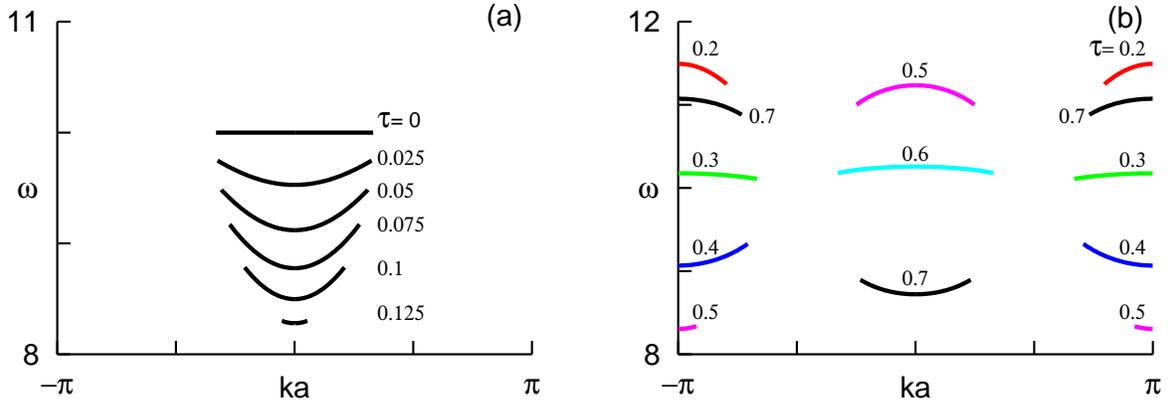}}}
\caption{Dispersion relation between allowed wave numbers and the corresponding
frequency shown as $\tau$ is gradually increased. $K=1$ and $\omega_0=10$.
A range of $\tau$ values is forbidden.
\label{FIG:wka_fortau}}
\end{figure}

Apart from these special cases, the 
general existence regions are complicated functions of
$ka$, $K$ and $\tau$. They need to be determined numerically by simultaneous solution
of equations~(\ref{EQN:freq}) and (\ref{EQN:amp}). To demonstrate the constraint imposed
by (\ref{EQN:freq}) we have plotted its solution ($\omega \;\;vs.\;\; ka$) for various values
of $\tau$ and for a fixed value of $\omega_0$ and $K$ in Fig.\ \ref{FIG:wka_fortau}(a). 
When $\tau=0$, the allowed range of $\tau$ 
is given by abs($ka$)$<\cos^{-1}(1-1/2K)$. So for $K=1$, the phase locked
patterns that have wave numbers less than $\pi/3$ are allowed,
and all of them have an identical frequency. As $\tau$ is increased
the frequency of oscillation decreases for small $\tau$, and 
the dispersion relation acquires a nonlinear parabolic character.
As $\tau$ is further increased, depending on the actual value of 
$K$, there are bands in $\tau$ values where no modes exist. 
The shrinking and disappearance of the dispersion curve at $ka=0$ 
beyond $\tau=0.125$ up to $\tau=0.2$ in Fig.\ \ref{FIG:wka_fortau}(a)-(b) illustrate this
phenomenon. 
One also notices from Fig.\ \ref{FIG:wka_fortau}(b) that at higher values of $\tau$
the dispersion curves become discontinuous and have bands of forbidden
$ka$ regions.


\section{Stability of phaselocked solutions}
\label{SEC:stab}

We now find the stability of
the equilibrium phaselocked solutions discussed in the previous section
by carrying out a linear perturbation analysis.  
Let,
\begin{equation}
\label{ansatzA}
\psi_j\left(t\right)=\left[R_ke^{i\omega_kt}
+u_j\left(t\right)\right]e^{i\left(jka\right)},
\end{equation}
where \(k=0,1,\cdots,N-1\). 
Substitution of (\ref{ansatzA}) in Eq.(\ref{EQN:ring}), yields in the lowest order 
the dispersion relation discussed in the previous section. In the next order,
where we retain terms that are linear in the perturbation amplitude, we 
observe that
\begin{equation}
\label{expand}
\mid\psi_k\mid^2\psi_k=\cdots+\left[
2R_k^2u_j\left(t\right)
+R_k^2e^{2i\omega_kt}\bar{u}_j\left(t\right)
\right]
e^{i\left(jka\right)}+\cdots.
\end{equation}
Using the above, we obtain the equation
\begin{equation}
\label{p1}
\begin{array}{rl}
\displaystyle{\frac{\partial u_j\left(t\right)}{\partial t}}
&=\ \ \left(1+i\omega_0-2R_k^2-2K\right)u_j\left(t\right)
-R_k^2e^{2i\omega_k t}\bar{u}_j\left(t\right)  \\
\ & \ \\
&\ \ \ +K\left[u_{j+1}\left(t-\tau\right)e^{ika}+
        u_{j-1}\left(t-\tau\right)e^{-ika}\right]
\end{array}
\end{equation}
and, taking its complex conjugate,
\begin{equation}
\label{p2}
\begin{array}{rl}
\displaystyle{\frac{\partial\bar{u}_j\left(t\right)}{\partial t}}
&=\ \ \left(1-i\omega_0-2R_k^2-2K\right)\bar{u}_j\left(t\right)
-R_k^2e^{-2i\omega_k t}u_j\left(t\right) \\
\ & \ \\
&\ \ \ +K\left[\bar{u}_{j+1}\left(t-\tau\right)e^{-ika}+
        \bar{u}_{j-1}\left(t-\tau\right)e^{ika}\right].
\end{array}
\end{equation}
Multiply Eqs.~(\ref{p1}) and (\ref{p2}) term-by-term by
\(e^{i\left(jqa\right)}\), make use of the identities
\begin{equation}
u_{j\pm 1}\left(t-\tau\right)e^{\pm ika}e^{i\left(jqa\right)}=
u_{j\pm 1}\left(t-\tau\right)e^{i\left(j\pm 1\right)qa}
e^{\pm i\left(k-q\right)a}
\end{equation}
and
\begin{equation}
\bar{u}_{j\pm 1}\left(t-\tau\right)e^{\mp ika}e^{i\left(jqa\right)}=
\bar{u}_{j\pm 1}\left(t-\tau\right)e^{i\left(j\pm 1\right)qa}
e^{\mp i\left(k+q\right)a},
\end{equation}
and sum over \(j=0,1,2,\cdots,N-1\). Introducing adjoint amplitudes
\(w_q\left(t\right)\) and \(\tilde{w}_q\left(t\right)\) by the
definitions
\begin{equation}
\left[w_q\left(t\right),\tilde{w}_q\left(t\right)\right] =
\sum_{j=0}^{N-1}\left[u_j\left(t\right),\bar{u}_j\left(t\right)
\right]e^{i\left(jqa\right)},
\end{equation}
we obtain the set of coupled equations
\begin{equation}
\begin{array}{rl}
\displaystyle{\frac{d w_q\left(t\right)}{dt}}
&=\ \ \left(1+i\omega_0-2R_k^2-2K\right)w_q\left(t\right)
-R_k^2e^{2i\omega_k t}\tilde{w}_q\left(t\right) \\
\ & \ \\
&\ \ \ +2K\cos\left[\left(k-q\right)a\right]
w_q\left(t-\tau\right)
\end{array}
\end{equation}
and
\begin{equation}
\begin{array}{rl}
\displaystyle{\frac{d\tilde{w}_q\left(t\right)}{dt}}
&=\ \ \left(1-i\omega_0-2R_k^2-2K\right)\tilde{w}_q\left(t\right)
-R_k^2e^{-2i\omega_k t}w_q\left(t\right) \\
\ & \ \\
&\ \ \ +2K\cos\left[\left(k+q\right)a\right]
\tilde{w}_q\left(t-\tau\right).
\end{array}
\end{equation}
In order to perform the stability analysis, we assume solutions of
the form
\begin{equation}
\left[w_q\left(t\right),\tilde{w}_q\left(t\right)\right]
=\left[ce^{i\omega_kt},\tilde{c}e^{-i\omega_kt}\right]
e^{\lambda t},
\end{equation}
which yield the set of coupled equations
\begin{eqnarray}
\label{con1}
C_+c+R_k^2\tilde{c}&=&0, \\
\label{con2}
R_k^2c+C_-\tilde{c}&=&0.
\end{eqnarray}
In these equations, the quantities \(C_{\pm}\) are given by
\begin{equation}
C_{\pm}=\lambda-\left[1\pm i\left(\omega_0-\omega_k\right)
-2R_k^2-2K\right]
-2K\cos\left[\left(k\mp q\right)a\right]
e^{\mp i\omega_k\tau}
e^{-\lambda\tau}.
\end{equation}
The eigenvalue equation is obtained from the determinantal
condition for Eqs.~(\ref{con1}) and (\ref{con2}), namely,
\begin{equation}
C_+C_--R_k^4=0.
\end{equation}
This can be expanded to be written in the form of the 
following characteristic equation,
\begin{equation}
\lambda^2 + (a_1 +  a_2) \lambda + (a_1 a_2 - R^4) = 0,
\label{EQN:eveqtau}
\end{equation}
where 
$a_1 = 2R^2 -1 + 2K -i (\omega_0-\omega)- 2K \cos[(k-q)a] {\mathrm e}^{-(\lambda+i\omega)\tau}$,
$a_2 = 2R^2 -1 + 2K +i (\omega_0-\omega)- 2K \cos[(k+q)a] {\mathrm e}^{-(\lambda-i\omega)\tau}$.

It should be noted that the perturbation wave numbers $q$ are once again a discrete set 
and from the periodicity requirement they obey the relation,
$$qa = m \frac{2\pi}{N}, ~~~m=0, 1, \ldots, N-1.$$
Thus in our stability analysis
any pattern corresponding to a given value of $ka$, we need to examine the eigenvalues of 
Eq.~(\ref{EQN:eveqtau}) at each of the above permitted values of $qa$. We now proceed to
discuss the stability of the various phaselocked patterns both in
the absence and presence of time delay.

\subsection{Stability of phaselocked patterns in the absence of delay}
In the absence of time delay, the eigenvalue equation (\ref{EQN:eveqtau})
can be solved analytically to give,
\begin{equation}
\lambda = 1-2R^2- 2 K + 2K \cos(ka) \cos(qa) \pm \sqrt{4 K^2\sin^2(ka)\sin^2(qa)+R^4 }.
\label{EQN:eval}
\end{equation}
The real parts of the eigenvalues of Eq.~(\ref{EQN:eveqtau}) will be negative if
$a_1+a_2>0$ and $a_1 a_2 - R^4 > 0$ simultaneously. 
The first of these conditions can be simplified to give
\begin{equation}
a_1+a_2 = 2\left[1- 2K \left\{ 1-\cos(ka) [2-\cos(qa)]\right\} \right] > 0, 
\label{EQN:co1}
\end{equation}
and the second condition can be simplified to give
\begin{equation}
a_1 a_2 - R^4 = 
[1-\cos(qa)] 4K \left[K\left\{4 \cos^2(ka)-2\cos(ka)-[1+\cos(qa)]\right\} + \cos(ka) \right] > 0.
\label{EQN:co2}
\end{equation}
In the following, we use these two conditions, or,
in the simplest cases, the eigenvalue equation itself to determine the
stability.

\subsubsection{In-phase patterns ($k=0$ mode)}
From the previous section the $k=0$ solution is given by 
$\omega = \omega_0$ and $R=1$.
Substituting these in Eq.~(\ref{EQN:eval}) above, and setting $k=0$, we get,
$$
\lambda = \cases{-2K [1-\cos(qa)]  \le 0,\cr
                 -2 - 2K[1-\cos(qa)] < 0,}
$$
Thus the plane wave solution $\Psi_j(t) = \mathrm{e}^{i\omega_0 t}$, which is 
nothing but an in-phase locked solution of the coupled identical oscillators,
is stable for $\tau=0$ for all values of $\omega_0$.

\subsubsection{Anti-phase patterns ($ka=\pi$ mode)}
The equilibrium pattern in this case consists of 
adjacent oscillators remaining $\pi$ 
out-of-phase at all times and oscillating with the same frequency $\omega_0$. 
The amplitude of the oscillations is given by, 
$$ R^2 = 1  - 4 K.$$
As discussed in the previous section, the condition $R^2 > 0$ permits the existence
of this mode in the region of $K<1/4$. However as we will see below, 
using conditions (\ref{EQN:co1}) and (\ref{EQN:co2}), 
these permitted anti-phase states are linearly unstable for any arbitrary number of oscillators.
Inserting $ka=\pi$ in (\ref{EQN:co1}) and (\ref{EQN:co2}) the conditions simplify to,
\begin{equation}
K < \mathrm{min}\frac{1}{2[3-\cos(qa)]},
\label{cond1}
\end{equation}
and
\begin{equation}
K > \mathrm{max}\frac{1}{5-\cos(qa)}.
\label{cond2}
\end{equation}
where the $qa$ values are to be determined as per the prescription discussed in
the previous section.  As simple examples, for $N=2$ the only permitted values
of $qa$ are $0$ and $\pi$ and from (\ref{cond1}) and (\ref{cond2}) above, we arrive at
the requirement that $K<1/8$ and $K>1/4$ which is not possible. Hence the mode
is unstable. Such conditions were previously obtained by Aronson {\it et al.}
\cite{AEK:90} in their investigation of the collective states of two coupled
limit cycle oscillators.
In fact, the argument can be extended to any arbitrary even $N$ since
$$
5-\cos(qa) \le 2[3-\cos(qa)]
$$
for all values of $qa$ and hence the conditions (\ref{cond1}) and (\ref{cond2})
cannot be satisfied simultaneously for any value of $K$.  This implies that the
anti-phase states, characterized by $ka=\pi$, are unstable for any arbitrary
value of even $N$.

\begin{figure}
\centerline{\rotatebox{270}{\includegraphics{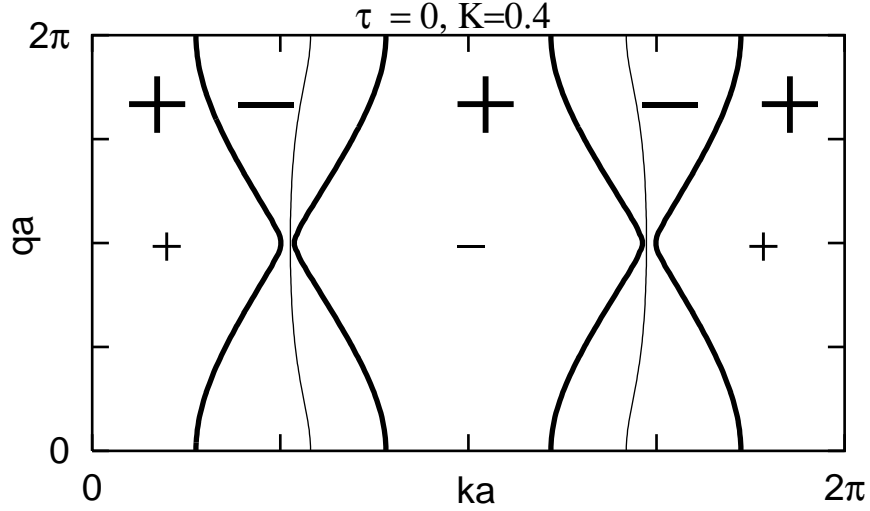}}}
\caption{The contours of $a_1+a_2=0$ (thin lines), and 
that of $a_1 a_2-R^4=0$ (thick lines). The thin $+$ and $-$ signs 
indicate the sign of $a_1+a_2$ in the connected regions bounded by
the thin curves, and the thick signs
indicate the sign of $a_1 a_2 - R^4$ in those bounded by
the thick curves.
\label{FIG:multipletau0plot}}
\end{figure}

\begin{figure}
\centerline{\rotatebox{270}{\includegraphics{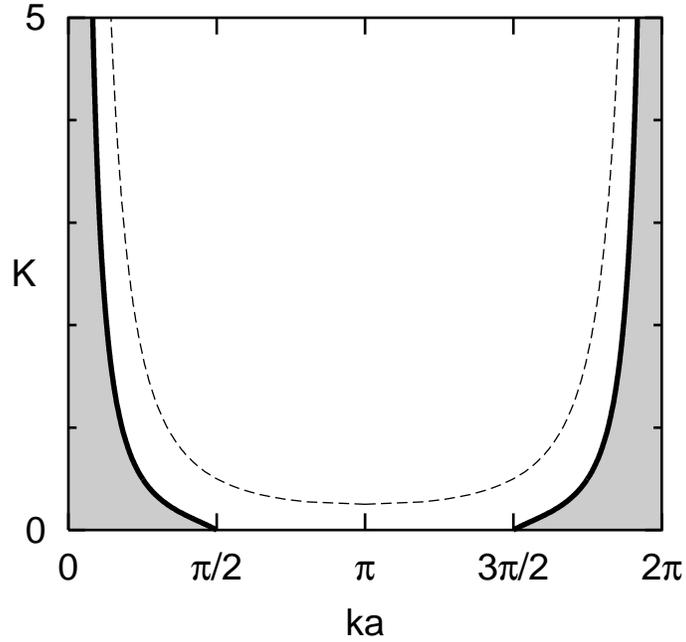}}}
\caption{The stability region of the phaselocked states (shaded region which
is below $K=K^\star$) plotted in $(K, ka)$ space. The left half of the dashed curve
is $f_1$ and the right half is $f_2$ as drawn in Fig.~\ref{FIG:forbiddentau}(a).
\label{FIG:stabkka}}
\end{figure}

\begin{figure}
\centerline{\scalebox{0.8}{\includegraphics{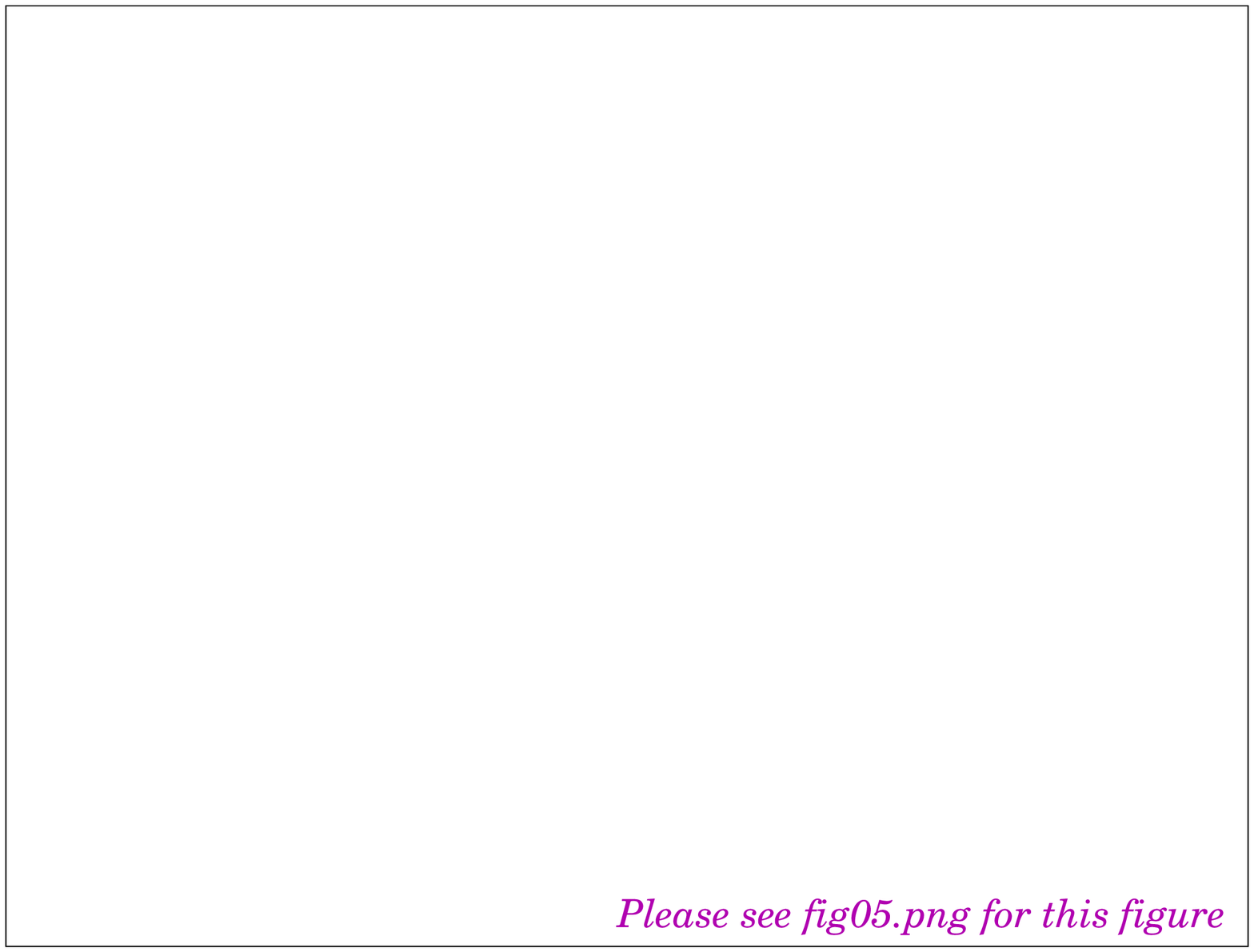}}}
\caption{Multiple stable phaselocked patterns for $N=50$ at $\omega_0 = 10$ and $K=0.4$ in the absence
of time delay. The real parts of the complex vector $Z_j(t)$ are plotted
in gray scale. \label{FIG:multipletau0}}
\end{figure}

\subsubsection{Other phaselocked patterns ($k>0$)}
Another interesting phaselocked pattern is the $k=\pi/2$ mode. 
It can be shown by using the conditions discussed above that this pattern
is also always unstable. However, a large number of other modes whose wave numbers are
close to either $0$ or $2\pi$ are likely to be stable. They coexist with
the in-phase stable solutions ($ka=0$). 
The relations (\ref{EQN:co1}) and (\ref{EQN:co2}) provide sufficient conditions
to find the stability of any given mode. The contours of $a_1+a_2=0$, 
and $a_1 a_2 - R^4 = 0$, are given, respectively, by
$$
qa = \left\{\cos^{-1}[2-(2K-1)/2KC], 2\pi-\cos^{-1}[2-(2K-1)/2KC] \right\},  
$$
$$
qa = \left\{\cos^{-1}\left[C(1/K-2+4C)-1\right],  2\pi-\cos^{-1}\left[C(1/K-2+4C)-1\right]\right\}, 
$$
where $C=\cos(ka)$.
These curves are plotted in Fig.~\ref{FIG:multipletau0plot} for $K=0.4$. We see that
there is a large range of $ka$ values that satisfies the conditions (\ref{EQN:co1}) 
and (\ref{EQN:co2}). For a given number $N$ of oscillators if the $ka$ values
fall in this range then the corresponding phaselocked patterns are stable.
The x-axis intersections of the curve $a_1 a_2 - R^4 = 0$ that are
closer to $0$ and $2\pi$ provide boundaries of $K$ below which
the modes are stable.  By setting $qa=0$ in (\ref{EQN:co2}), we obtain
the condition on $K$ for stability as a function of $ka$:
\begin{equation}
K < \frac{-\cos(ka)}{2[2\cos^2(ka)-\cos(ka)-1]} \equiv K^\star .
\end{equation}
This stability region is plotted in Fig.~\ref{FIG:stabkka}.  In the limit of
$N\rightarrow\infty$, a continuous range of $ka$ modes are accessible, and the
system truly possesses infinitely many stable phaselocked states when
$K<K^\star$. The stability of these phaselocked solutions is one important
result of our paper. Note that all phaselocked patterns with wave numbers
between $\pi/2$ and $3\pi/2$ are unstable.  We also note from
Fig.~\ref{FIG:stabkka} that for a given value of $K$ it is possible to have
more than one stable state corresponding to different values of $ka$ that lie
in the two stable regions. In the limit of $N \rightarrow \infty$ such a
multistability phenomenon can occur over a continuous range of $ka$ values
spanning the stable regions.  A numerical example of the multistability of some
of these modes is illustrated in Fig.~\ref{FIG:multipletau0} for $N=50$
oscillators at a fixed value of $K=0.4$.  By giving initial conditions close to
the modes $ka= 2\pi/50$, $4\pi/50$, and $6\pi/50$, the corresponding
phaselocked solutions are realized.

Now we address the following question: What is the minimum number of
oscillators for which a second phaselocked pattern (the first one being
in-phase) can become stable? Since the permissible $ka$ values are related to
the number of oscillators $N$ in an inverse fashion due to the periodicity
requirement (see Eq.~(\ref{spect})) we need to determine the maximum stable $ka$ in
order to find the minimum critical $N$.  Since $K=K^\star$ makes an
intersection at $ka=\pi/2$ and $ka=3\pi/2$, the second stable phaselocked
pattern that could emerge at this point will have 

$$ 
ka = m \frac{2\pi}{N} = \frac{\pi}{2}.
$$
Hence $N\ge 4$ for a second phaselocked pattern to ever become stable.
Thus for  $N=2$ or $N=3$ coupled identical oscillators we will only have
in-phase oscillations as stable oscillations and there will not be 
any other stable phaselocked states. The minimum number of oscillators necessary for
multistable behavior is four. Thus $N=4$ is a critical number.

\subsection{Phaselocked patterns for finite time delay}
\label{SEC:lockdelay}

Now we study the synchronized patterns that could become stabilized by
the presence of time delay. 
We noted before that, in the absence of time delay, 
the in-phase oscillations are always stable, and the anti-phase oscillations
are always unstable for any positive finite $K$.
Time delay changes the scenario and could make each of these branches 
stable in certain ranges of $\tau$. In addition, the number of multiple
branches of each of these two oscillations increases with $\tau$.
Equations (\ref{EQN:freq}-\ref{EQN:amp})
define the amplitudes and frequencies of these synchronized states. 
and the characteristic equation (\ref{EQN:eveqtau}) determines their linear
stability.

\begin{figure}
\centerline{\rotatebox{270}{\scalebox{0.8}{\includegraphics{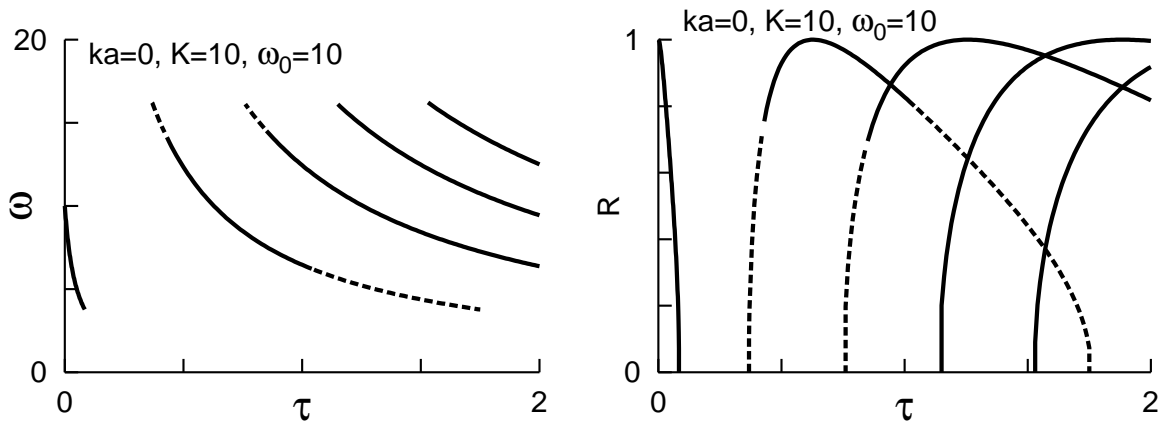}}}}
\caption{The in-phase ($ka=0$) locked frequencies and 
amplitudes represented by Eq.~(\ref{EQN:ring})
at $K=10$, and $\omega_0=10$. The dashed portions represent the unstable branches
for the case of $N=2$ and the continuous lines represent the stable branches.
\label{FIG:wrtau_ka0_fortau}}
\end{figure}

\begin{figure}
\centerline{\rotatebox{270}{\scalebox{0.8}{\includegraphics{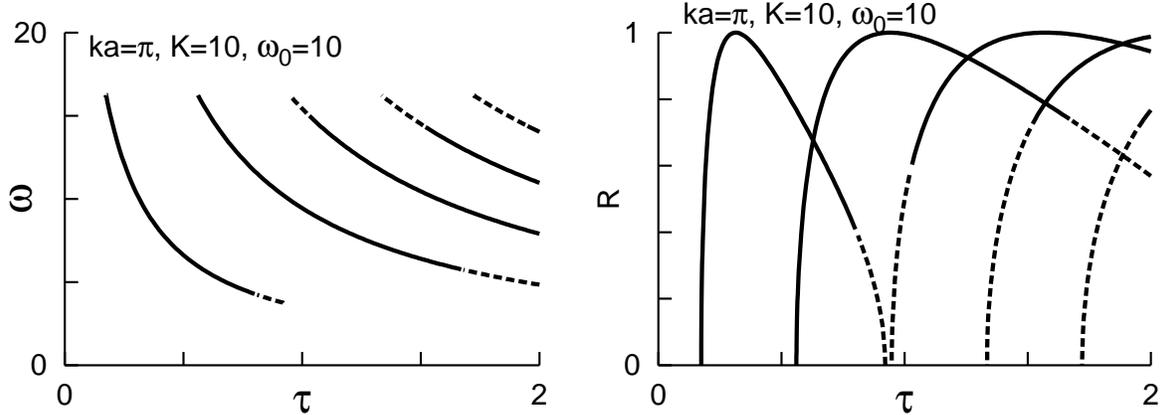}}}}
\caption{The anti-phase ($ka=\pi$) locked frequencies and 
amplitudes represented by Eq.~(\ref{EQN:ring})
at $K=10$, and $\omega_0=10$. The dashed portions represent the unstable branches
for the case of $N=2$ and the continuous lines represent the stable branches.
\label{FIG:wrtau_kapi_fortau}}
\end{figure}

We begin with a simple example of two coupled oscillators, $N=2$, which can
have just two phaselocked states, namely in-phase states with $ka=0$ and
anti-phase states with $ka=\pi$.  We evaluate the eigenvalues for these two
cases for different values of $\tau$ by numerically solving
Eq.~(\ref{EQN:eveqtau}) for the permitted values of $qa$ and also confirm the
existence of the modes by an actual integration of the original equations. We
show the stable (continuous) and unstable (dashed lines) branches of these two
solutions in Figs.\ \ref{FIG:wrtau_ka0_fortau} and \ref{FIG:wrtau_kapi_fortau}.
A stable in-phase branch emerges from
$\tau = 0$ and its amplitude becomes $0$ while its frequency is finite
showing a supercritical Hopf bifurcation. The first branch of the
anti-phase locked state emerges from zero again in a supercritical
Hopf. As $\tau$ is increased multiple Hopf bifurcation points are
seen. Such stability of the in-phase and anti-phase oscillations were
earlier studied by us experimentally using coupled electronic
oscillators \cite{RSJ:00exp}.

\begin{figure}
\centerline{\rotatebox{270}{\scalebox{1.0}{\includegraphics{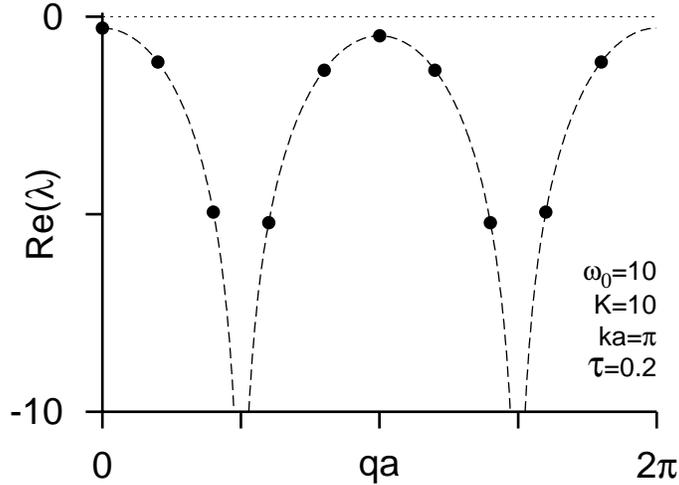}}}}
\caption{Real part of the eigenvalues as a function of $qa$ for $N=10$ at $\omega_0=10$, 
$K=10$, $ka=\pi$, and $\tau=0.2$.
\label{FIG:relvsqan10}}
\end{figure}

\begin{figure}
\centerline{\rotatebox{270}{\scalebox{0.8}{\includegraphics{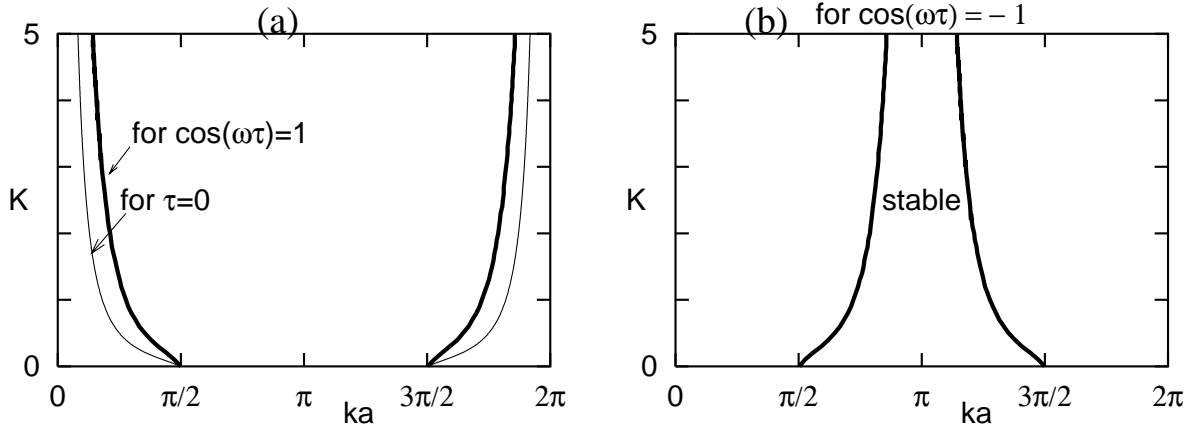}}}}
\caption{(a) Critical boundaries below which stable phaselocked patterns exist
are shown for the case of $\cos(\omega\tau) = 1$ as $N\to\infty$.
(b) Stability region of phaselocked patterns for $\cos(\omega\tau)=-1$ as $N\to\infty$.
\label{FIG:critKka_tau}}
\end{figure}

\begin{figure}
\centerline{\includegraphics{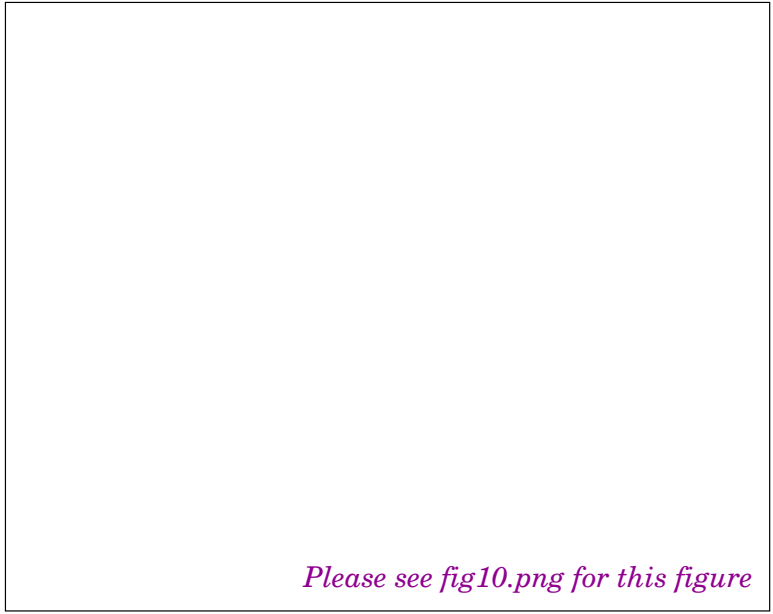}}
\caption{Anti-phase locked state of identical oscillators for $N=50$ at $K=10$ and $\tau=0.3$.
\label{FIG:kpitau}}
\end{figure}

As a second illustration, we consider the case of $N=10$, $\tau=\pi/\omega_0$
(i.e.  $\cos(\omega \tau) =-1.0$), whose equilibrium existence domain we have
discussed earlier as case (iv) in Section \ref{SEC:disp}. 
In particular we have noted that the $ka=\pi$ state is a
permitted equilibrium state. To examine its stability we observe that we need
to examine the eigenvalues for $qa = m \frac{2\pi}{N}, ~~~m=0, 1, \ldots, N-1$.
In Fig.~\ref{FIG:relvsqan10} we have plotted the real part of the eigenvalues
for all these perturbation wave numbers. We see that they are all negative,
which indicates stability of the pattern. In fact, in this case, due to the
symmetry of the characteristic equation we can predict stability for all higher
values of $N$ as well with the eigenvalues of all additional $qa$ values
falling on the dotted line shown in the figure. The existence regions for this
case were shown in Fig.\ \ref{FIG:forbiddentau}(d). By considering this case
for all the possible values of $qa$ between $0$ and $2\pi$, we are essentially
determining the stability of this case in the infinite oscillator limit. We
show our numerical results for this case in Fig.\ \ref{FIG:critKka_tau}. It is
obtained by tracking the eigenvalue transitions for $qa$ between $0$ and
$2\pi$. This phase diagram is strictly true for $N\to\infty$. As can be seen,
finite $\tau$ increases the number of stable modes for any given value of $K$.
We finally show in Fig. \ref{FIG:kpitau} a numerical example of the anti-phase
oscillations using $N=50$ oscillators.

\section{Amplitude death for finite time delay}
\label{SEC:death}

As is well known, a system of identical oscillators that are globally coupled
do not have an amplitude death state \cite{AEK:90} but the presence of time
delay in the coupling can bring about such a collective state. This phenomenon
was first pointed out for two coupled oscillators in \cite{RSJ:98} and also
generalized to $N$ globally coupled oscillators \cite{RSJ:99}. A similar
statement also holds for locally coupled oscillators as will be shown. The
death state in the presence of time delay had earlier been numerically
confirmed for a finite number of locally coupled oscillators \cite{RSJ:98}.
However no systematic study of the dependence of the death island regions on
the magnitude of the time delay and the number of oscillators has so far been
carried out for systems of locally coupled oscillators. In this section we
address this issue and study the stability of the origin for our discrete model
equation (\ref{EQN:ring}). Amplitude death state is characterized by $R=0$.  To
derive an appropriate set of eigenvalue equations for determining the stability
of this state, we carry out a linear perturbation analysis about the origin.
Substituting ($ \psi_{j} = 0 + \tilde{\psi}_{j}$) in Eq.\ (\ref{EQN:ring}), and
discarding the nonlinearities, we get,
\begin{equation}
\frac{\partial \tilde{\psi}_{j}}{\partial t} 
= (1 + i \omega_0 )\tilde{\psi}_{j} 
+ K [ \tilde{\psi}_{j+1}(t-\tau) - \tilde{\psi}_{j}(t) ] 
+ K [ \tilde{\psi}_{j-1}(t-\tau) - \tilde{\psi}_{j}(t) ], 
\end{equation}
where $j=1, \ldots, N$ with periodic boundary conditions.
By letting $\tilde{\psi}(t) \propto \mathrm{e}^{i\lambda t}$, 
the eigenvalue matrix in circulant form is obtained.
The determinant of this matrix is written as
$$
\prod_{j=1}^{N}\left (\lambda +2K -1 -i\omega_{0} 
-K e^{-\lambda \tau}U_{j} 
-K e^{-\lambda \tau}U_{j}^{N-1}\right )= 0,
$$
where $U_{j} = \mathrm{e}^{i2\pi(j-1)/N}$ are the $N^{th}$ roots of unity.
But $U_j + U_j^{N-1}= U_j+U_j^{-1} = 2\cos[(j-1)2\pi/N]$.
So the above equation takes the form of
\begin{equation}
\prod_{j=1}^{N}\left (\lambda +2K -1 -i\omega_{0} -2K \cos[(j-1)2\pi/N]
\mathrm{e}^{-\lambda \tau}\right) = 0 .
\label{EQN:evaltau}
\end{equation}
The complete set of eigenvalue equations includes the second set obtained by
considering the conjugate equation of the above. Note that for $\tau=0$ the
above eigenvalue equation (\ref{EQN:evaltau}) always admits at least one
unstable eigenvalue, namely $\lambda = 1 +i\omega_{0}$. Hence identical
oscillators that are locally coupled cannot have an amplitude death state in
the absence of time delay.  We will now determine the amplitude death regions
for finite values of $\tau$.  We define a factor $R_j = 2\cos[(j-1)2\pi/N]$
that we will use in the critical curves derived below.  If the number of
oscillators is a multiple of $4$, there are some eigenvalue equations that
emerge without a dependence on $\tau$, when $R_j=0$. For example, consider the
case of $N=4$ and $j=2,4$.  Then, the eigenvalue equation becomes: $\lambda = 1
- 2 K \pm i \omega_0$.  For this equation, the only criticality is given by
$K=1/2$.  The stable region lies on the side of the parameter space that obeys
$K>1/2$. For other values of $R_j$, the death island boundaries can be derived
by setting the real part of the eigenvalue to zero, and appropriately choosing
the signs of the multiple curves that result. The analysis is similar to the
one we presented in the treatment of globally coupled oscillators
\cite{RSJ:98,RSJ:99}, and here we simply provide the final expressions for the
critical curves in $(\tau,K)$ plane: 
\begin{eqnarray}
\tau_a(n,K)= \cases{%
\frac{2n\pi-\cos^{-1}[(2K-1)/KR_j]}{\omega_0+\sqrt{K^2R_j^2-(2K-1)^2}}, ~~~ R_j>0, \cr
\frac{(2n+1)\pi-\cos^{-1}[(2K-1)/K\left|R_j\right|]}{\omega_0+\sqrt{K^2R_j^2-(2K-1)^2}}, ~~~ R_j<0,}
\label{EQN:taua}
\end{eqnarray}
\begin{eqnarray}
\tau_b(m,K)=\cases{
\frac{2m\pi+\cos^{-1}[(2K-1)/KR_j]}{\omega_0-\sqrt{K^2R_j^2-(2K-1)^2}}, ~~~R_j>0,\cr
\frac{(2m+1)\pi+\cos^{-1}[(2K-1)/K\left|R_j\right|]}{\omega_0-\sqrt{K^2R_j^2-(2K-1)^2}}, ~~~R_j<0 .
}
\label{EQN:taub}
\end{eqnarray}
$n$ and $m$ are whole numbers.  We now determine some useful bounds on $K$ for
ordering and finding the degeneracies of the critical curves.  The argument of
inverse cosine functions and the square root term in the denominators of the
above expressions impose the following bounds on $K$:
\begin{eqnarray}
K > 1/4, ~~\mathrm{for}~~ R_{j=1},  \label{EQN:bound1}\\
\frac{1}{2+R_j} < K < \frac{1}{2-R_j}, ~~\mathrm{for}~~R_{j>1}. \label{EQN:bound2}
\end{eqnarray}
The sign of the derivative with respect to $\tau$ of the real part of the
eigenvalue is determined by the term $-\mathrm{Im}(\lambda) K R_j
\sin[\mathrm{Im}(\lambda) \tau]$, and after some algebra it is expressed as
\begin{equation}
\frac{d\mathrm{Re}(\lambda)}{d\tau}\Bigg|_{\mathrm{Re}(\lambda)=0} \cases{%
> 0, ~~\mathrm{on}~~ \tau_a, ~~\mathrm{for~any}~\omega_0, \cr
< 0, ~~\mathrm{on}~~ \tau_b, ~~\mathrm{if}~~\omega_0 > \sqrt{K^2 R_j^2 - (2K-1)^2},\cr
> 0, ~~\mathrm{on}~~ \tau_b, ~~\mathrm{if}~~\omega_0 < \sqrt{K^2 R_j^2 - (2K-1)^2}.
}
\end{equation}
The first condition involving $\omega_0$ in the above
relation leads to the following bounds on $K$:
\begin{eqnarray}
K<(1+\omega_0^2)/4, ~~\mathrm{for}~~R_{j=1}, \\ \label{EQN:bound3}
K > \frac{2+R_j\sqrt{1+\omega_0^2}}{4-R_j^2}, ~~\mathrm{or,}~~
K < \frac{2-R_j\sqrt{1+\omega_0^2}}{4-R_j^2}, ~~\mathrm{for}~~R_{j>1}. \label{EQN:bound4}
\end{eqnarray}
The ranges of $K$ imposed by the relations (\ref{EQN:bound1}) and
(\ref{EQN:bound3}) are mutually exclusive.  Hence, on the curve $\tau_b$ the
only boundary across which an eigenvalue pair makes a transition to the
negative eigenvalue plane occurs when $R_{j=1}$, and across the curves
occurring when $R_{j>1}$, the eigenvalue pair makes a transition to the
positive plane.  So $\tau_b(0,K)$ at $R_{j=1}$ always remains as the left hand
side boundary of the death island.  In order to see the degeneracy among the
curves, note that $$R_k=R_j, ~~\mathrm{if ~~}k=N+2-j.$$ Also the sign of $R_j$
does not play a role in distinguishing the critical curves. These two
properties are responsible for a reduction of the number of the actual distinct
boundaries of the death islands. These degeneracies can be framed into two
cases.  First, when $N$ is odd. The total number of distinct values of $R_j$ is
$(N+1)/2.$ $R_1 (=2)$ is the maximum of all the $R_j$, and is non-degenerate.
And $R_i$, where $i=2, \ldots, (N+1)/2$ are the other distinct values, whose
values are identically equal to $R_k$, where $k=N+2-i$.  Among the latter $R_j$
values, the maximum negative value occurs at $j=(N+1)/2$. The curves
corresponding to $R_j>0$ and $R_j<0$ form the boundaries for two different
death islands, as can be seen from the indexes in
Eqs.~(\ref{EQN:taua}-\ref{EQN:taub}).  A further ordering of the curves must be
done by a numerical plotting of the curves. The ordering reveals that for $N
\le 13$, the first death island is bounded by the curves 
\begin{eqnarray}
\gamma_1 &: & ~\tau_b(0,K) ~~\mathrm{at}~~ R_{j=1}, \nonumber \\
\gamma_2 &: & ~\tau_b(0,K) ~~\mathrm{at}~~ R_{j=(N+1)/2}, ~~\mathrm{and}~~ \nonumber \\
\delta_1 &: & ~\tau_a(0,K) ~~\mathrm{at}~~ R_{j=(N+1)/2}, \nonumber \\
\delta_2 &: & ~\tau_a(1,K) ~~\mathrm{at}~~ R_{j=1}. \nonumber
\end{eqnarray}
The curves $\gamma_1$ and $\delta_2$ form, respectively, the left and the right
boundaries of the death island region.  And $\delta_1$ and $\delta_2$ form the
bottom two curves. The existence range along $K$ of $\delta_1$, across which
the eigenvalues make a transition to the left half plane, increases with
increasing $N$. In fact it intersects with $\gamma_1$. So this provides the
first boundary across which stability is lost. This occurs for $N\ge 15$, when
the death island is bounded by the two curves:
\begin{eqnarray}
\gamma_1 &: & ~\tau_b(0,K) ~~\mathrm{at}~~ R_{j=1},  ~~\mathrm{and}  \nonumber\\
\delta_1 &: & \tau_a(0,K) ~~\mathrm{at}~~ R_{j=(N+1)/2}. \nonumber
\end{eqnarray}

\begin{figure}
\centerline{\rotatebox{270}{\includegraphics{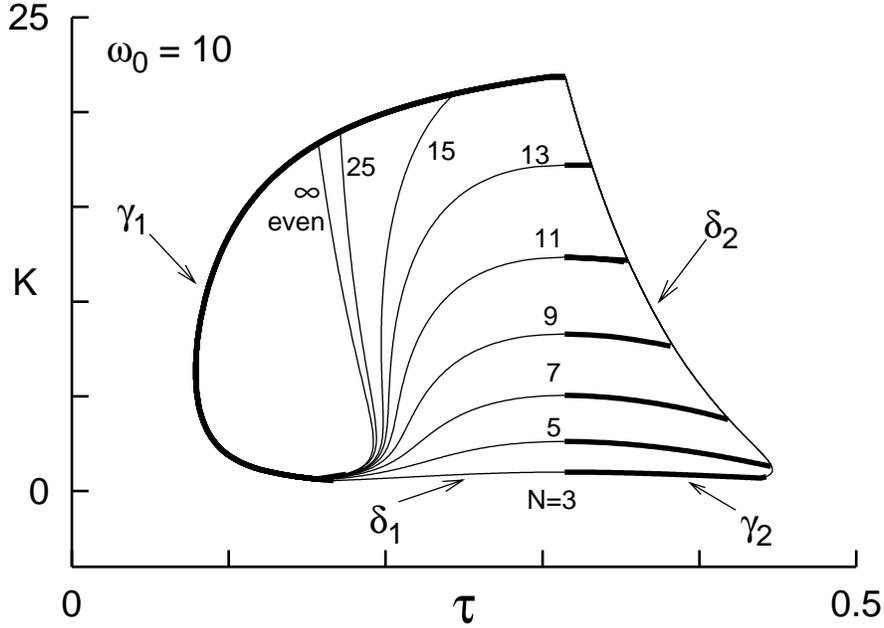}}}
\caption{Death islands at $\omega_0=10$.
All even number of oscillators have a single death island region
that is independent of the number of oscillators. The odd number of 
oscillators are bounded by four curves when $N\le 13$, and two curves
otherwise. These two curves merge in the infinite limit with the
curves that represent the even number of oscillators.
\label{FIG:dislnds}}
\end{figure}

\begin{figure}
\centerline{\includegraphics{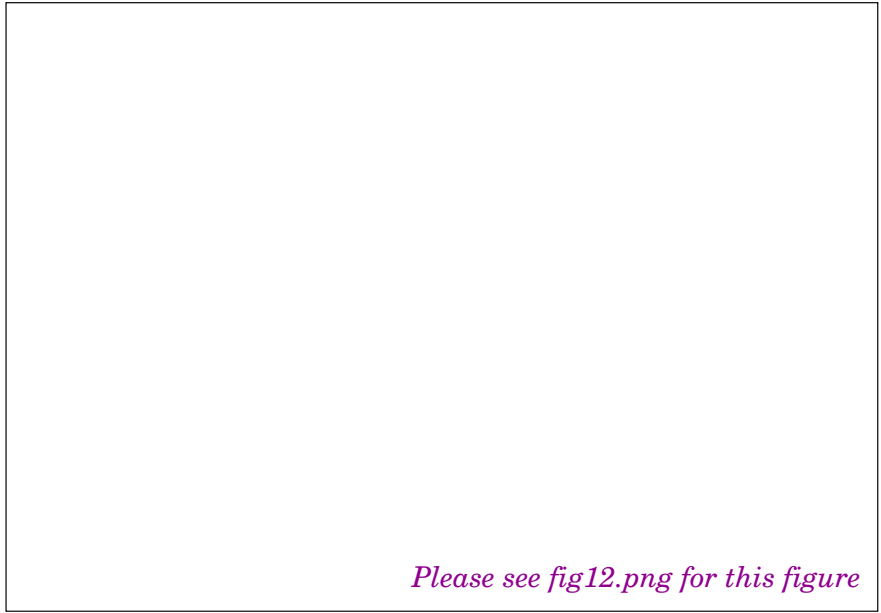}}
\caption{Amplitude death of identical oscillators for $N=50$ at $K=10$ and $\tau=0.1$.
\label{FIG:death}}
\end{figure}

Second, when $N$ is even, many more curves become degenerate.  The number of
distinct value of $R_j$ is $N/2+1.$ Among these distinct values, the magnitudes
of pairs of them can become identical.  If $N$ is divisible by $4$, such pairs
are $N/4+1$ in number, and $(N+2)/4$ otherwise. Every positive $R_j$ has its
negative counterpart.  The maximum value of $R_j = 2$ that occurs when $j=1$,
and the next maximum value is at $R_{j=2}$. Hence the equations
(\ref{EQN:taua}-\ref{EQN:taub}) can be simplified to
\begin{eqnarray}
\tau_c(n,K)= 
\frac{n\pi-\cos^{-1}[(2K-1)/KR_j]}{\omega_0+\sqrt{K^2R_j^2-(2K-1)^2}}, ~~~ R_{j=1,\ldots,J},
\label{EQN:tauaeven} \\
\tau_d(m,K)= \frac{m\pi+\cos^{-1}[(2K-1)/KR_j]}{\omega_0-\sqrt{K^2R_j^2-(2K-1)^2}}, ~~~R_{j=1,\ldots,J},
\label{EQN:taubeven}
\end{eqnarray}
where $J=N/4+1$ if $N$ is divisible by $4$, and $(N+2)/4$ otherwise.  The
actual ordering, however, reveals that the death island boundaries are given by
$\tau_c(0,K)$ at $R_{j=1}$ and $\tau_d(1,K)$ at $R_{j=1}$, which are identical,
respectively, to $\delta_1$ and $\gamma_1$.  These curves are plotted in Fig.\
\ref{FIG:dislnds}. As is seen for $N$ even there is a single death region. In
the case of $N$ odd the boundary of the death region depends on the value of
$N$. As $N$ increases, the area of the death region decreases. As $N
\rightarrow \infty$ the area of the death island for $N$ odd decreases and
approaches, as a limit, the boundary for $N$ even.  For all $N$, the
intersections of $\gamma_1$ and $\delta_1$ or those of $\gamma_2$ and
$\delta_2$ occur for $K>1/2$. So the delay-independent eigenvalue equations
that are mentioned earlier do not contribute to the death island boundaries.
The differences in the death island boundaries for even and odd numbered
oscillators can be traced primarily to the behavior of the eigenvalues of the
lowest permitted perturbation wave numbers. For an even number $N$ of
oscillators the smallest perturbation mode is $qa=\pi$. The values of the real
parts of the eigenvalues corresponding to this mode are close in their
magnitude to those corresponding to the $qa=0$ perturbation mode.  Across the
right hand side boundary of the death island region, the $qa=\pi$ mode grows
positive and the system emerges out of the death region with an anti-phase
state. When $N$ is odd, however, the smallest perturbation mode is
$qa=\pi-\pi/N$ which is more heavily damped than the $qa=0$ mode. So the death
region continues to exist for larger $\tau$ values. Ultimately, the second
eigenvalue branch of the $qa=0$ mode (which exists due to the transcendental
nature of the eigenvalue equation) grows and the system emerges across the
boundary with an in-phase state of a different frequency.  As $N$ becomes
large, the smallest perturbation mode for the $N$ odd case gets closer to $\pi$
and the death island boundaries of the two cases, as seen in Fig.\ \ref{FIG:dislnds}, 
become indistinguishable.  We have independently verified
the death regions depicted in Fig.\ \ref{FIG:dislnds}, including their
interesting dependence on $N$, by a direct numerical solution of
(\ref{EQN:ring}) over the specified range of parameters.  We also illustrate in
Fig.~\ref{FIG:death} an example of the amplitude death state for a chain of
$N=50$ oscillators.

\section{Discussion and conclusions}
\label{SEC:con}

We have studied the existence and stability of phaselocked patterns and
amplitude death states in a closed chain of delay coupled identical limit cycle
oscillators that are near a supercritical Hopf bifurcation.  The coupling is
limited to nearest neighbors and is linear. The coupled oscillators are modeled
by a set of discrete dynamical equations. Using the method of plane waves we
have analyzed these equations and obtained a general dispersion relation to
delineate the existence regions of equilibrium phaselocked patterns. We have
also studied the stability of these states by carrying out a linear
perturbation analysis around their equilibria.  Our principal results are in
the form of analytic expressions that are valid for an arbitrary number of
oscillators $N$ (including the $N \rightarrow \infty$ thermodynamic limit) and
that can be used in a convenient fashion to identify and or obtain stable
equilibrium states for a given set of parametric values of time delay, coupling
constant, wave number and wave frequency. We have carried out such an exercise
for a number of illustrative cases both with and without the presence of time
delay.  In the absence of time delay, our analysis reveals a number of new
phase-locked states close to the in-phase stable state which can exist
simultaneously with the in-phase state. The minimum number of oscillators for
which this multirhythmic phenomena can occur is $N=4$. Time delay introduces
interesting new features in the equilibrium and stability scenario. In general
we have found that time delay expands the range of possible phaselocked
patterns and also extends the stability region relative to the case of no time
delay. The dispersion curves for varying values of time delay also display some
novel features such as forbidden regions and jumps in the range of allowed wave
numbers as well as forbidden bands in the space of time delay.

We have also carried out a detailed analytic and numerical investigation of the
existence of stable amplitude death states in the closed chain of delay coupled
identical limit cycle oscillators. The results not only confirm our earlier
numerical demonstration of the existence of such death states but go beyond to
provide a comprehensive picture of the existence regions in the parameter space
of time delay and coupling strength. The analytic results also establish that
death island regions exist for any number of oscillators $N$ for appropriate
vales of $K$, $\omega$ and $\tau$.  In this sense our work provides a
generalization of the earlier amplitude death related results, that were
obtained for globally coupled oscillators to the case of locally coupled
oscillators. An interesting new result, arising from the local coupling
configuration, is that the size of a {\it death island} is independent of $N$
when $N$ is even but is a decreasing function of $N$ when $N$ is odd. In other
words the death island results for the $N=2$ island hold good for any arbitrary
even number of locally coupled oscillators and constitute the minimum size of
the death island in the $K\;-\;\tau$ parameter space. This can have interesting
practical implications. For example in coupled magnetron or laser applications
if one is seeking to minimize the parametric region where death may occur (and
thereby greatly diminish the total power output of the system) it is best to
select a configuration with an even number of devices. At a more fundamental
level this ``invariance'' property which is strongly dependent on the symmetry
of the system may also have interesting dynamical consequences, e.g. in the
manner of the collective relaxation toward the death state from arbitrary
initial conditions. In fact we see some evidence of this differing dynamics in
our numerical investigations of the time evolution of the system toward the
death state. For an even number of oscillators we observe a rapid clustering of
the oscillators into two distinct groups that are $\pi$ out of phase. These two
giant clusters then slowly pull each other off their orbits and spiral toward
the origin.  For an odd number of oscillators the lack of symmetry appears to
prevent this grouping, the phase distribution of the oscillators has a greater
spread in its distribution and the relaxation dynamics is distinctly different.
As the number of oscillators increases and the asymmetry gets reduced the
difference in the dynamical behavior becomes less distinct. In the limit of $N
\rightarrow \infty$, the difference vanishes and the size and shape of the odd
$N$ island asymptotes to the $N$-even island. A fundamental understanding of
this dynamical behavior and its relation to the symmetry dependence emerging
from the stability analysis could be an interesting area of future exploration.
Our results could also be useful in applications where locally coupled
configurations are employed such as in coupled magnetron devices, coupled laser
systems, and neural networks as a roadmap for accessing their various
collective states.


\end{document}